\documentclass[aps,prd,amsmath,amssymb,
twocolumn,
floatfix,
superscriptaddress,
]{revtex4-1}
\usepackage{graphicx}
\usepackage{times}
\usepackage{multirow}
\usepackage{verbatim}
\usepackage{color,soul}
\usepackage{xcolor}
\usepackage{cancel}
\usepackage[normalem]{ulem}

\graphicspath{{figures/}}

\newcommand{\uFive}{$^{235}$U}
\newcommand{\uEight}{$^{238}$U}
\newcommand{\pNine}{$^{239}$Pu}
\newcommand{\pOne}{$^{241}$Pu}

\begin{document}


\title{Impact of Fission Neutron Energies on Reactor Antineutrino Spectra}

\author{B. R. Littlejohn}%
 \email{blittlej@iit.edu}
\affiliation{Physics Department, Illinois Institute of Technology, Chicago, IL 60616, USA}%

\author{A. Conant}%
\affiliation{Department of Nuclear and Radiological Engineering, Georgia Tech, Atlanta, GA 30332, USA}%

\author{D. A. Dwyer}%
\affiliation{Lawrence Berkeley National Laboratory, Berkeley, CA 94720, USA}%

\author{A. Erickson}%
\affiliation{Department of Nuclear and Radiological Engineering, Georgia Tech, Atlanta, GA 30332, USA}%

\author{I. Gustafson}%
\affiliation{Physics Department, Illinois Institute of Technology, Chicago, IL 60616, USA}%

\author{K. Hermanek}%
\affiliation{Physics Department, Illinois Institute of Technology, Chicago, IL 60616, USA}%

\begin{abstract}
Recent measurements of reactor-produced antineutrino fluxes and energy spectra are inconsistent with models based on measured thermal fission beta spectra.
In this paper, we examine the dependence of antineutrino production on fission neutron energy.  
In particular, the variation of fission product yields with neutron energy has been considered as a possible source of the discrepancies between antineutrino observations and models.
In simulations of low-enriched and highly-enriched reactor core designs, we find a substantial fraction of fissions  (from 5\% to more than 40\%) are caused by non-thermal neutrons.  
Using tabulated evaluations of nuclear fission and decay, we estimate the variation in antineutrino emission by the prominent fission parents $^{235}$U, $^{239}$Pu, and $^{241}$Pu versus neutron energy.  
The differences in fission neutron energy are found to produce less than 1\% variation in detected antineutrino rate per fission of $^{235}$U, $^{239}$Pu, and $^{241}$Pu.  
Corresponding variations in the antineutrino spectrum are found to be less than 10\% below 7~MeV antineutrino energy, smaller than current model uncertainties.  
We conclude that insufficient modeling of fission neutron energy is unlikely to be the cause of the various reactor anomalies.
Our results also suggest that comparisons of antineutrino measurements at low-enriched and highly-enriched reactors can safely neglect the differences in the distributions of their fission neutron energies.
\end{abstract}

\maketitle

\section{Introduction}

Numerous breakthroughs in particle physics have been achieved through measurement of reactor electron antineutrinos.  
Recent noteworthy developments~\cite{bib:prl_rate,bib:reno,bib:mention2011}, particularly the measurement of the neutrino mixing parameter $\theta_{13}$, have relied on relative comparisons of measured $\overline{\nu}_e$ fluxes at multiple locations.  
Direct comparison of measured reactor $\overline{\nu}_e$ fluxes to those predicted by reactor $\overline{\nu}_e$ models~\cite{bib:reinesCC, bib:reinesNC, KamLAND, bib:prl_reactor, reno_bump, bib:prl_evol} have also yielded important results, despite facing substantial model uncertainties.  
Direct model comparisons will continue to be of importance in near-term efforts to perform precision measurements of the neutrino mass hierarchy and Standard Model oscillation parameters at medium-baseline experiments~\cite{juno1} and efforts to probe the origin of various reactor antineutrino anomalies with short-baseline experiments~\cite{prospect,solid,stereo}.  

The favored method of predicting reactor $\overline{\nu}_e$ production is via conversion of  measured electron spectra to the corresponding antineutrino spectra emitted during beta decay of fission daughters.  
The primary inputs for this method are cumulative spectroscopic measurements of beta particles produced during the decay of daughters produced by thermal fission of actinides relevant for nuclear reactors: $^{235}$U, $^{239}$Pu, and $^{241}$Pu~\cite{bib:ILL_1,bib:ILL_2}.  
These beta spectrum measurements were performed in the 1980s at the Institut Laue-Langevin (ILL), by exposing foils of these actinides to thermal neutrons from a beamline emanating from ILL's High Flux Reactor.
The measured beta spectrum are fitted as the sum of a number of 'virtual' beta branches with representative nuclear charges, which are then kinematically converted into corresponding $\overline{\nu}_e$ spectra, taking into account a variety of second-order corrections~\cite{bib:ILL_3,Vogel}.  
The sum of the converted, corrected spectra for all virtual beta branches forms the predicted $\overline{\nu}_e$ spectrum for each fissioning isotope.  
The beta spectrum following $^{238}$U fission was not measured at ILL's thermal neutron beamline, since the vast majority of $^{238}$U~fissions are induced by fast neutrons.  
This isotope contributes less than 10\% of all fissions in reactors of interest to antineutrino experiments.
For this isotope, flux predictions have been formed via \textit{ab initio} (or `summation') calculations utilizing standard fission yield and beta decay measurement databases~\cite{bib:mueller2011}; it should be noted that beta-conversion $^{238}$U $\overline{\nu}_e$ spectra above 2.875~MeV have been recently calculated utilizing new fast-neutron-induced fission beta spectrum measurements~\cite{bib:munich}.

The beta-conversion $\overline{\nu}_e$ $^{235}$U, $^{239}$Pu, and $^{241}$Pu spectra of Huber~\cite{bib:huber} are currently the most widely-used in the field, in part due to the high precision and well-defined uncertainties that have been associated with the conversion procedure.  
Summation predictions are in principle less precise, with $\sim$10\% associated uncertainties in their absolute normalizations and comparatively less well-defined spectrum shape uncertainties.  
Despite these limitations, summation predictions are a valuable tool for assessing systematic uncertainties in reactor $\overline{\nu}_e$ predictions, as demonstrated previously in Refs.~\cite{bib:HayesShape,bib:dwyer,bib:sonzogni,sonzongi2}.  

Increasing precision in both $\overline{\nu}_e$ flux and spectrum measurements have uncovered a variety of inconsistencies with recent beta-conversion models. 
A series of measurements dating back to the early 1980s at a variety of reactors show a $\sim$6\% deficit in detected reactor $\overline{\nu}_e$-induced inverse beta decays (IBDs) with respect to the recent beta-conversion reactor model~\cite{bib:mention2011}; this discrepancy has been termed the `reactor antineutrino flux anomaly.'
The reactor $\theta_{13}$ experiments have validated the existence of this IBD yield deficit~\cite{bib:prl_reactor,bib:reno_2016}, while also uncovering discrepancies with respect to the beta-conversion prediction's spectrum shape, in particular a $\sim$10\% detected excess in the 5-7 MeV region of $\overline{\nu}_e$ energy~\cite{bib:prl_reactor,bib:cpc_reactor,reno_bump}; the latter discrepancy has been referred to as the `reactor spectrum anomaly.'  
More recent Daya Bay results indicate that the size of the measured $\sim$6\% IBD yield deficit is dependent on the content of nearby reactor cores, indicating that incorrect flux predictions are at least partially responsible for the reactor antineutrino flux anomaly~\cite{bib:prl_evol}.  

Significant discussion in the nuclear and particle physics community has focused on diagnosing possible issues with existing beta-conversion predictions.  
The oscillation of reactor $\overline{\nu}_e$ to sterile neutrinos is one popular hypothesis that questions underlying particle physics models, as opposed to nuclear physics models~\cite{bib:mention2011, bib:giuntiGlobal}.  
This hypothesis fits existing hints from other neutrino experiments~\cite{anomalywhite}, but cannot fully explain the reactor spectrum or flux anomalies.  
Other hypotheses relate more directly to the formulating elements of the beta-conversion predictions to which reactor $\overline{\nu}_e$ data are compared.  
A variety of studies have examined in detail the impact of differing treatment of the forbiddenness~\cite{bib:hayes,hayesEvol} or nuclear charge~\cite{hayesEvol,bib:HayesShape} of virtual beta branches, as well as the impact of uncertainties in the various second-order conversion corrections~\cite{bib:HayesZ,bib:HayesMag}.  
Others have questioned the accuracy of the underlying fission beta spectrum measurements~\cite{bib:dwyer,hayes2} or have suggested~\cite{hayes2,haser} or ruled out~\cite{bib:HayesShape} issues with specific parent fission isotopes.  

\begin{figure}[htb!pb]
\includegraphics[trim=0.3cm 0.1cm 2.0cm 2.0cm, clip=true, width=0.49\textwidth]{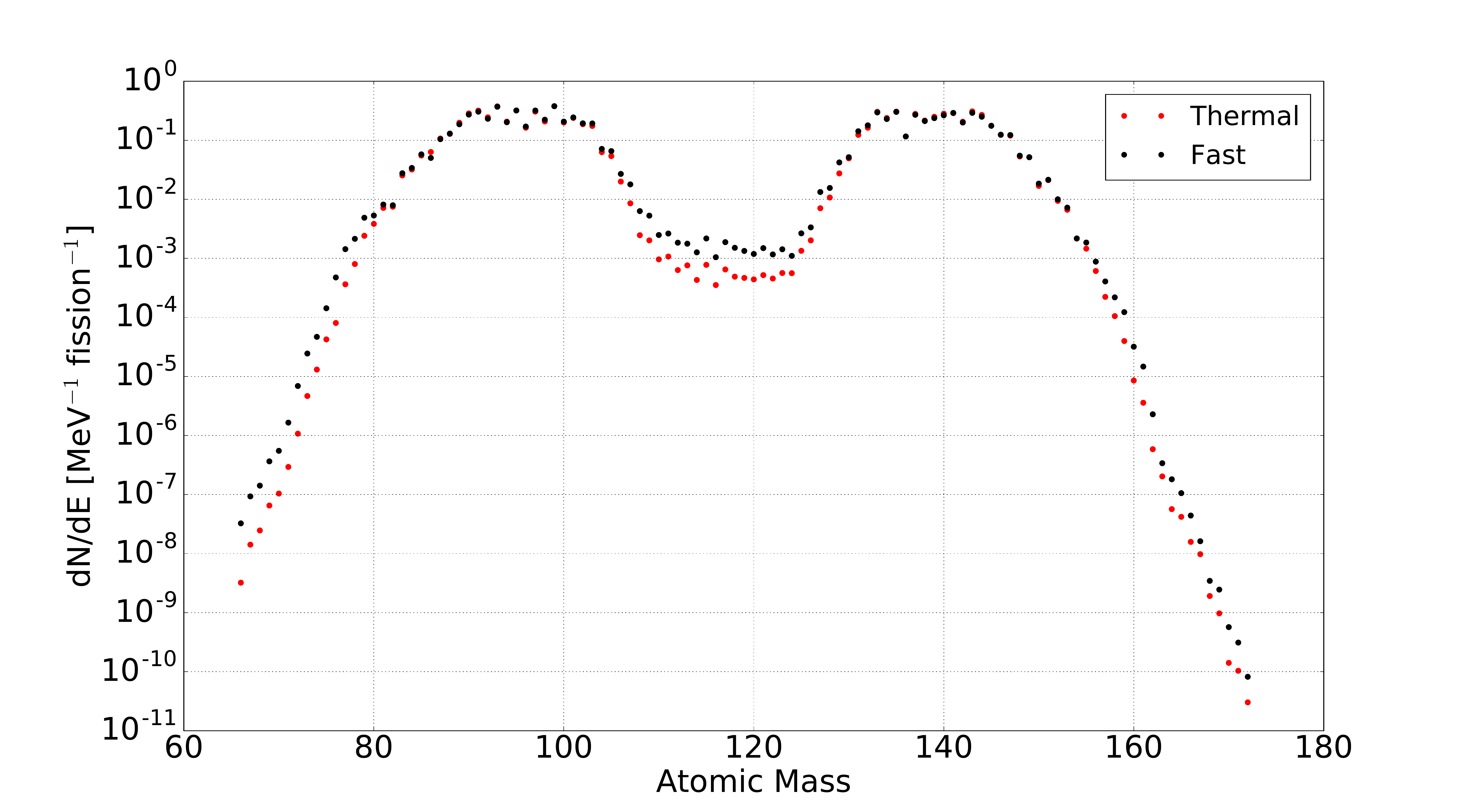}
\caption{Cumulative fission product yields for~$^{235}$U versus mass number for fission induced by thermal (red) and fast (black) neutrons, as given by the JEFF-3.1.1 database.  An increase in yield from fast neutrons is visible in the valley between the maximum yield peaks.}
\label{fig:yields}
\end{figure}

It has also been suggested that differing fission neutron energies between $\overline{\nu}_e$ and beta spectrum measurements may be to blame for the observed rate and spectrum discrepancies~\cite{hayes2}.  
A substantial number of $\overline{\nu}_e$ generated in reactor cores are products of resonant or fast fission, while in ILL's thermal neutron beamline, measured betas were generated only by products of thermal-induced fission.  
This is an important consideration, as fission product yields are well-known to be dependent on incident neutron energy.  
An example of this is provided in Figure~\ref{fig:yields}, using fission yields from the JEFF-3.1.1 database~\cite{bib:JEFF}: yields of products in the valley between the two peaks of the mass-yield curve are known to increase with incident neutron energy.  
A complete picture of neutron-energy-related impacts on $\overline{\nu}_e$ production isotopes has not been well-defined in the literature.   
Experimental measurements of fission product yields at specific neutron energies are challenging with respect to other common fission-related measurements, such as variations in kinetic energy release and neutron multiplicity.  
Prior studies have provided indications of substantial $>$10\%-level differences in thermal and epithermal fission yields for specific isotopes and neutron energy ranges~\cite{Popa,Tong}.  
Meanwhile, some more recent work examining deviations from thermal fission mass-yield curves over the entire epithermal range \cite{Leconte2012} or at a select few resonant energies \cite{Gook2017} provide indications of only percent-level variations.  
Theoretical predictions of energy-dependent fission product yields are also challenging, and currently do not provide a usefully complete picture.
With these limitations in mind, it is not surprising that the impact of non-thermal fissions is not considered in existing beta-conversion predictions.

However, a variety of scenarios involving non-thermal fissions could meaningfully impact the comparison of reactor-produced $\overline{\nu}_e$ measurements and ILL's beta-conversion $\overline{\nu}_e$ predictions:
\begin{itemize}
\item{Reduced relative production of isotopes with Q-values above the 1.8~MeV IBD threshold by resonant and fast fissions could result in suppressed production of $\overline{\nu}_e$ in reactors -- and reduced IBD yields in nearby $\overline{\nu}_e$ detectors -- relative to thermal neutron beamlines.  This scenario could produce the observed reactor antineutrino flux anomaly.}
\item{Increased relative production of isotopes with Q-values above 5~MeV by resonant and fast fissions could result in an excess of detected IBDs at high energies at conventional reactors.  This scenario could produce the observed reactor spectrum anomaly.}
\item{Increased or decreased relative production of isotopes with Q-values above 1.8~MeV by resonant and fast fissions from \textit{only select parent fission isotopes} could result in discrepancies in the evolution of IBD yields between Daya Bay and beta-conversion predictions.}
\end{itemize}

\begin{table*}[t!]
  \caption{Relative contributions to total fissions for three intervals of neutron energy for the four prominent fissioning isotopes in a PWR reactor.  Values are reported for three periods: beginning-, middle-, and end-of-cycle (BOC,MOC,EOC). The percentages for each isotope sum to unity for each period.}
    \begin{tabular}{l|ccc|ccc|ccc}
    \hline \hline
          &  & \textit{Thermal} & & & \textit{Resonance} & & & \textit{Fast} &  \\
          & BOC & MOC & EOC & BOC & MOC & EOC & BOC & MOC & EOC \\
          \hline
    $^{235}$U & 80.86\% & 82.43\% & 83.12\% & 16.68\% & 15.34\% & 14.76\% & 2.46\% & 2.23\% & 2.13\% \\
    $^{238}$U & 0.00\% & 0.00\% & 0.00\% & 0.06\% & 0.06\% & 0.06\% & 99.94\% & 99.94\% & 99.94\% \\
    $^{239}$Pu & 27.27\% & 29.66\% & 30.88\% & 71.72\% & 69.44\% & 68.24\% & 1.01\% & 0.90\% & 0.87\% \\
    $^{241}$Pu & 35.03\% & 37.79\% & 38.83\% & 64.05\% & 61.42\% & 60.40\% & 0.92\% & 0.80\% & 0.77\% \\
    \hline \hline
    \end{tabular}%
  \label{tab:pwr_energy_point1ev}%
\end{table*}%

The current study estimates the size of neutron-energy-related effects on reactor $\overline{\nu}_e$ production and detection.
We use the summation method to compare predictions of $\overline{\nu}_e$ fluxes for a variety of reactor scenarios with a particular focus on fission neutron energy and its impact on fission yields.
First, reactor core simulations are used to estimate the relative actinide fission rates versus neutron energy for three types of reactors.
Next, we combine these reactor simulations with the expected variation in fission daughter yields versus neutron energy, as given in the JEFF and ENDF nuclear databases, to estimate the corresponding change in $\overline{\nu}_e$ production.
For a low-enriched uranium (LEU) and two highly-enriched uranium (HEU) core types, the resultant $\overline{\nu}_e$ flux and spectrum is then compared to the case of purely thermal neutron-induced fission, as well as among these three reactor core types.
Using this method, we then investigate the three scenarios outlined above, which, if present, could explain the observed IBD yield and spectrum discrepancies relative to beta-conversion predictions.  


In Section~\ref{sec:ReacSim} of the paper, we describe the reactor simulations used in this study, and summarize the key findings relevant to fission neutron energies.
In Section~\ref{sec:SpecTools}, we describe the nuclear databases and how these are used to generate reactor $\overline{\nu}_e$ spectra for different fission neutron energy distributions.  
Generated spectra are then presented and discussed in Section~\ref{sec:SpecPresent}, with limitations and additional cross-checks discussed in Section~\ref{sec:Syst}.  
The implications of this study are then summarized in Section~\ref{sec:Implications}.

\section{Reactor Simulations} 
\label{sec:ReacSim}


Reactor simulations of several different reactor types were performed using MCNP~\cite{bib:mcnp}, a frequently used Monte Carlo transport code with capabilities in nuclear reactor simulations. MCNP5 was used due to extensive benchmarking of all MCNP5-based models used in this study. The list of reactor models investigated for this study are: a typical Westinghouse pressurized water reactor (PWR)~\cite{LWR_MOX}, the High Flux Isotope Reactor (HFIR) at Oak Ridge National Laboratory (ORNL)~\cite{Ilas15}, and the National Bureau of Standards Reactor (NBSR) at the National Institute of Standards and Technology (NIST)~\cite{NBSR}.  
The Westinghouse PWR operates with LEU fuel, while HFIR and NBSR operate with HEU fuel.  

The pressurized water reactor (PWR) is a type of standard commercial power reactor that operates in the United States. The relatively large core, approximately 3-4 meters in both diameter and active length, generates a power of approximately 3500 MWt. The fuel is in the form of UO$_2$ cylindrical fuel pellets grouped into fuel rods. A 17-by-17 square lattice of fuel rods constitute fuel assemblies, which are placed in the core. The fresh fuel enrichment is approximately 4.2\%. Typical commercial reactors operate for 18-24 months prior to fuel shuffling and refueling. The PWR MCNP model~\cite{LWR_MOX} is an eight-core representation of a standard Westinghouse four-loop core. The model includes an accurate batch representation of fuel (fresh, once-burnt, and twice-burnt), and burnable absorbers as neutron poisons. 

The High Flux Isotope Reactor (HFIR) is a research reactor that is primarily used for neutron scattering experiments, materials irradiation, and neutron activation analysis. The relatively small core, on the order of 0.5 meters in both diameter and active fuel height, generates a power of approximately 85 MWt. The dispersion fuel is in the form of 93\% enriched U$_3$O$_8$-Al in thin, involute-shaped fuel plates. The fuel plates are grouped into two regions, the inner- and outer- fuel elements, which total to 540 plates in total in the HFIR core. HFIR operates for approximately 24 days, after which a shutdown allows for a new core replacement. The HFIR MCNP full-core model~\cite{Ilas15} includes explicit geometry of the involute fuel plates with separate radial and axial fuel regions.

The National Bureau of Standards Reactor (NBSR) is a research reactor used for purposes similar to those of HFIR. The core is also relatively small, approximately 1.12 meters in diameter and 0.74 meters in height. Its nominal thermal power is 20 MWt, about a quarter of that of HFIR. The fuel is also in the form of 93\% enriched U$_3$O$_8$-Al curved plates. There are two fuel sections, upper and lower, which are separated by an unfueled gap.

MCNP~\cite{bib:mcnp} modeling used ENDF-B-VII.0~\cite{bib:ENDF} for all nuclear reactions, including the major fission isotopes. MCNP uses a track-length estimator of each neutron with trajectory in a cell to calculate the energy-dependent neutron flux in the cell. Flux tallies were estimated for each cell in user-defined energy bins, which can be further manipulated, e.g., multiplied by fission cross sections, to obtain energy-dependent reaction rates. 

\begin{table*}[t!]
  \caption{Relative contributions to total fissions for three intervals of neutron energy for the four prominent fissioning isotopes in PWR, HFIR, and NBSR reactors. All values are reported for the middle-of-cycle, except for the NBSR where the values are for ~1.5 days into the cycle.}
\begin{tabular}{l|cccc|cccc|cccc}
    \hline \hline
          & \multicolumn{4}{c|}{\textit{Thermal}} & \multicolumn{4}{c|}{\textit{Resonance}} & \multicolumn{4}{c}{\textit{Fast}} \\
          & Beam & PWR & HFIR & NBSR & Beam & PWR & HFIR & NBSR & Beam & PWR & HFIR & NBSR \\
          \hline
    $^{235}$U  & 100.00\% & 82.43\% & 85.88\% & 94.38\% & 0.00\% & 15.34\% & 12.36\% & 4.93\% & 0.00\% & 2.23\% & 1.76\% & 0.69\% \\
    $^{238}$U  & NA & 0.00\% & NA & NA & NA & 0.05\% & NA & NA & NA & 99.94\% & NA & NA \\
    $^{239}$Pu & 100.00\% & 29.66\% & NA & NA & 0.00\% & 69.44\% & NA & NA & 0.00\% & 0.90\% & NA & NA \\
    $^{241}$Pu & 100.00\% & 37.79\% & NA & NA & 0.00\% & 61.42\% & NA & NA & 0.00\% & 0.80\% & NA & NA \\
    \hline \hline
    \end{tabular}%
  \label{tab:fissionenergy}%
\end{table*}%

In this study, we choose to focus on the fractional contribution of neutrons in specific energy ranges to all fissions in each core type.  
Fractional fission contributions are broken down into three major groups of fission neutron kinetic energy: thermal, resonant, and fast.  
Resonant neutrons, which exist in a region of dense neutron-induced fission cross-section resonances for the fission isotopes $^{235}$U,  $^{239}$Pu, and  $^{241}$Pu, are defined to be above 0.1 eV and below 10 keV for $^{239}$Pu and $^{241}$Pu, and above 1 eV and below 10 keV for $^{235}$U.  
The resonant definition is lowered for $^{239}$Pu and $^{241}$Pu due to the presence of a large cross-section resonance centered at 0.2-0.3~eV, which is not present in $^{235}$U.   
Fast and thermal neutrons are defined to be all energies above and below the resonant region, respectively.  
As a first example, Table~\ref{tab:pwr_energy_point1ev} summarizes the calculated fractional fission contribution of each neutron energy range for the PWR reactor model at beginning of cycle (BOC, Day 0), near the middle of the cycle (MOC, Day 350), and near the end of the 18-month fuel cycle (EOC, Day 560).  
Statistical uncertainties in these modeled values are less than 0.1\%.  
Errors due to underlying neutron interaction cross-section uncertainties are not considered in this analysis; ENDF covariance data for \uFive~and \pNine~fission cross-sections are under 0.5\% in the thermal region and under 10\% for nearly all resonances, including in the 0.2-0.3~eV resonance region described above for the plutonium isotopes.  
We do not expect uncertainties of this size to meaningfully alter the findings of this study.  

In this model, thermal neutrons account for 82.4\%, 29.7\%, and 37.8\% of all \uFive, \pNine, and \pOne~fissions, respectively, while fissions from neutrons in the resonance range account for 16.7\%, 69.4\%, and 61.4\%, respectively.  
These values deviate significantly from the 100\% thermal fraction experienced by the irradiated actinide foils in the ILL beta spectrum measurements.  
Due to low fission cross-sections at low neutron energies, nearly 100\% of all \uEight~fissions are induced by fast neutrons; for the other isotopes, fast fissions account for only 1-2\% or less of the total.  
Fractional contributions from each energy range are constant to within a few percentage points over the entire cycle, with most of this variation falling in the first few days of the cycle.


\begin{figure}[htb!pb]
\includegraphics[trim=0.3cm 0.1cm 1.0cm 1.0cm, clip=true, width=0.48\textwidth]{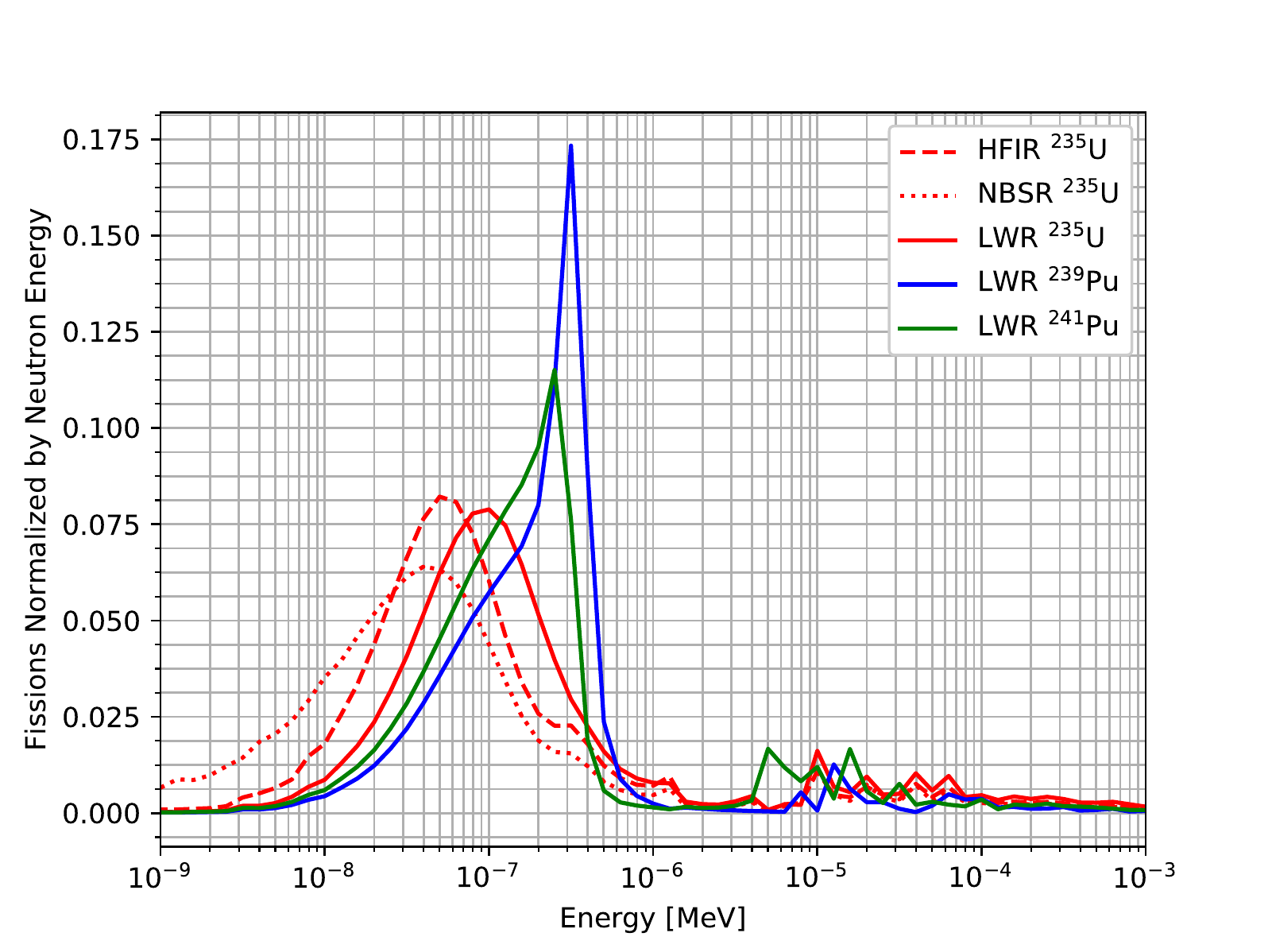}
\includegraphics[trim=0.3cm 0.1cm 1.0cm 1.2cm, clip=true, width=0.48\textwidth]{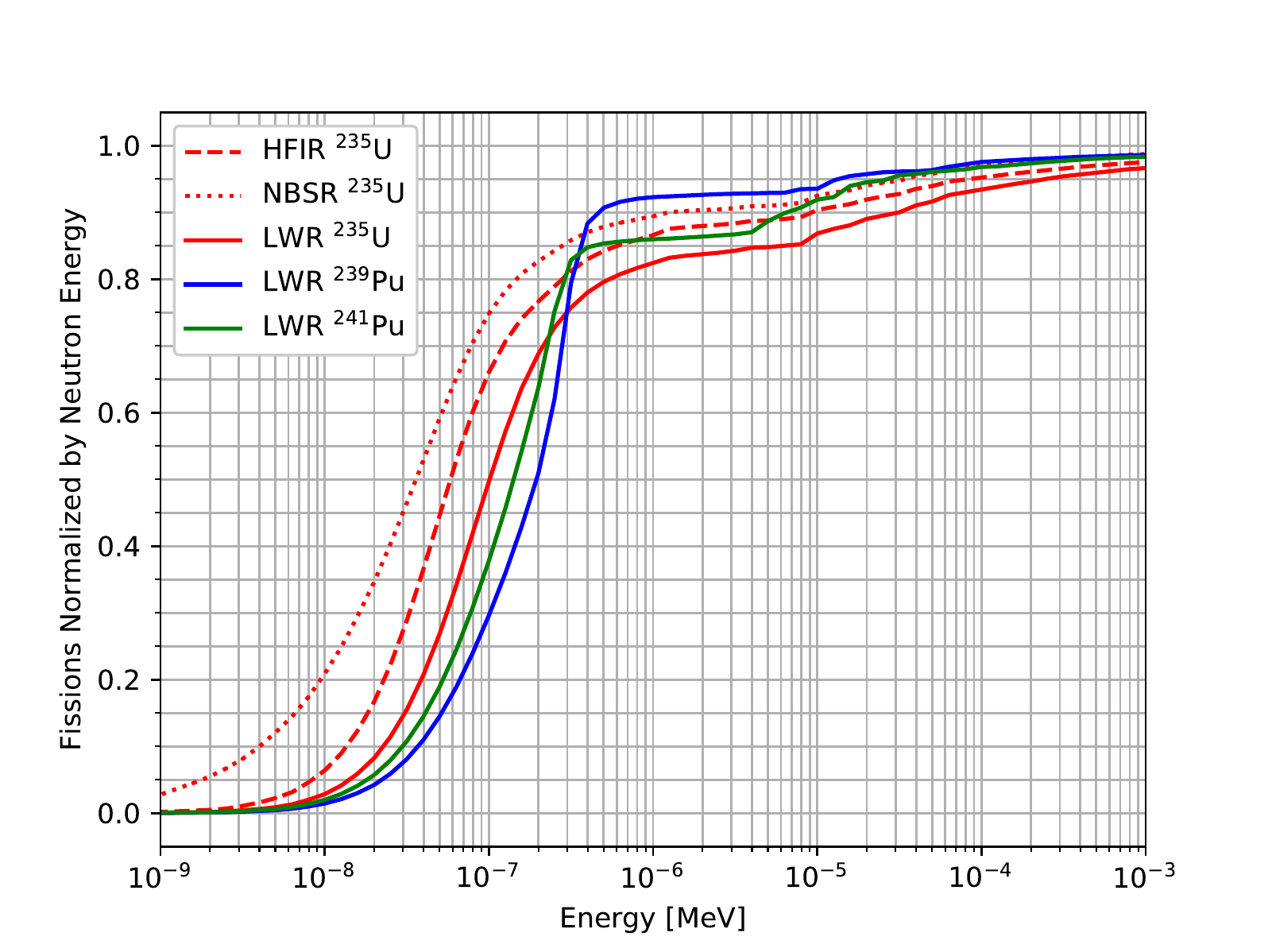}
\caption{Normalized (top) and cumulative (bottom) fission contributions versus incident neutron energy for $^{235}$U, $^{239}$Pu, or $^{241}$Pu in PWR/LWR, HFIR, and NBSR reactors. As HFIR and NBSR reactors are HEU, curves are only provided for $^{235}$U.}
\label{fig:neutronspectra}
\end{figure}


These substantial non-thermal fission rates are described in more detail in Figure~\ref{fig:neutronspectra}, which plots relative fission contribution versus incident neutron energy, as well as the cumulative contribution with increasing neutron energy.  
Peaks and discontinuities are visible across the resonant range for all three isotopes, mirroring underlying resonances in fission cross-sections in this region.  
The most striking peaks occur at 0.3 eV and 0.2~eV in \pNine~and \pOne, respectively, where a large and broad resonance appears in each isotope's fission cross-section, as previously mentioned.  
These two peaks visually dominate the Gaussian-shaped contribution in the thermal fission region, contributing 64\% and 50\% of all fissions for \pNine~and \pOne, respectively, at MOC.  
The remaining resonances at higher epithermal energies contribute only 9\% and 11\% of total fissions for these isotopes, respectively.  

The fractional fission contributions for different energy ranges are compared between reactor types in Figure~\ref{fig:neutronspectra} and in Table~\ref{tab:fissionenergy}.  
Looking at the \uFive~curves in Figure~\ref{fig:neutronspectra}, a substantial shift downward in fission neutron energy is visible in the thermal regime for the HFIR model compared to the PWR model; this difference mirrors the comparatively higher proportion of thermal neutrons in the HFIR core.  
This trend toward lower neutron energy is even more strongly exhibited in the NBSR HEU core model. Thus, it appears that the fission neutron energies of the HEU reactors appear to most closely resemble those of the pure-thermal ILL beta spectrum measurements.  

In subsequent sections, we will investigate whether these non-trivial differences in fission neutron energies between the ILL thermal neutron beamline, PWR reactors, and HEU reactors might give rise to measurable differences in their $\overline{\nu}_e$ spectra.  


\section{Description of the Summation Calculation and its Inputs}
\label{sec:SpecTools}

For a specific parent fission isotope, summation (or \textit{ab initio}) reactor $\overline{\nu}_e$ spectrum predictions are formed from the sum of beta spectra from all branches of all fission products~\cite{VogelHayesReview,bib:fallot}:
\begin{equation}\label{eq:sum1}
\frac{dS(E_{\nu})}{dt} = \sum_i R_i \sum_j f_{ij} S_{ij}(E_{\nu}),
\end{equation}
where $R_i$ is the decay rate of fission product $i$, $f_{ij}$ is the probability of decay (also called the branching fraction or beta feeding) to excited state $j$ of fission product $i$'s beta decay daughter, and $S_{ij}(E_{\nu})$ is the beta spectrum produced by this decay, which is dependent on the Q-value of the decay, the energy level of daughter excited state $j$, the spin and parity of the parent ground state and daughter excited state $j$, and a host of other corrections to the standard Fermi beta spectrum shape.  
For a reactor in equilibrium, the decay rate $R_i$ of short-lived isotopes, which dominate the overall $\overline{\nu}_e$ flux, will be equivalent to the overall fission rate, $R^f$, times the cumulative fission yield $Y^{c}_i$, or the probability that a fission will produce isotope $i$ either through direct production via fission or through subsequent beta decay of other fission products.  
In summation calculations, cumulative fission yields $Y^c_i$, beta feedings $f_{ij}$ and the properties of parent and daughter nuclei determining $S_{ij}(E_{\nu})$ are generally taken from one of a number of standard nuclear data tables.  
Fission rates $R^f$ can be either removed by considering the $\overline{\nu}_e$ flux and spectrum \textit{per fission}, or can be determined via simulation of the reactor core under consideration using one of a number of community-standard reactor simulation packages, as was described in Section~\ref{sec:ReacSim}.  

For the present analysis, we would also like to consider separately the rate of fissions induced by neutrons from each of the three primary energy categories: thermal, resonant and fast.  
In this case, Eq.~\ref{eq:sum1} must be altered to include an additional sum over neutron energy category:
\begin{equation}\label{eq:sum2}
\frac{dS(E_{\nu})}{dt} = \sum_{k=t,r,f} R^f_k \sum_i Y^c_{ik} \sum_j f_{ij} S_{ij}(E_{\nu}).
\end{equation}
As before, nuclear databases provide cumulative fission yields $Y^c_{ik}$, beta feedings $f_{ij}$, and $S_{ij}(E_{\nu})$, while reactor simulations can be used to determine $R^f_k$.  

For this study, the Python-based toolkit Oklo~\cite{bib:Oklo} was used to produce summation-predicted $\overline{\nu}_e$ fluxes and spectra.  
The Oklo toolkit takes as input the fission yield databases ENDF-B-VII.1~\cite{bib:ENDF}, JEFF-3.1.1, and JENDL-4.0~\cite{bib:JENDL} for $Y^c_{ik}$, and tabulations of ENSDF nuclear structure data files for $f_{ij}$ and $S_{ij}$~\cite{Anna}.  
In addition to producing and plotting beta and $\overline{\nu}_e$ predictions as described by Eq.~\ref{eq:sum2}, Oklo also includes functionality for easy parsing and analysis of the input databases, and for simple calculation and plotting of spectra for subsets of branches or fission products.  

All three fission yield databases contain separate entries for thermal and fast fissions, but not for resonant fissions.  
To overcome this limitation, we assume that resonant fission daughter yields are equivalent to those from fast fission.  
This approach should provide a conservative overestimate of the relative differences in $\overline{\nu}_e$ spectrum and flux between the pure thermal and non-thermal reactor cases. 

\section{Reactor Antineutrino Production Versus Fission Neutron Energy}
\label{sec:SpecPresent}

As a default case, we consider $\overline{\nu}_e$  spectra generated using JEFF-3.1.1 fission yields and beta decay information from the previously-mentioned ENSDF tabulation, supplemented by a tabulation of theoretically-predicted beta spectra for some isotopes with incomplete nuclear structure database information~\cite{Sonzogni}.   
Fission yields from JEFF are used to avoid apparent mistakes and limitations in some ENDF-reported fission yields~\cite{sonzongi2}, some of which are reproduced in the JENDL database. 
For thermal $^{235}$U fissions, 96\% of this default predicted $\overline{\nu}_e$ flux above the 1.8~MeV inverse beta decay interaction threshold is produced by ENSDF-specified isotopes; these daughter isotopes' $\overline{\nu}_e$ spectra $\sum_j S_{ij}(E_{\nu})$ are calculated as described above.  
The theoretically-calculated flux, which is provided directly as a function of energy for each isotope, accounts for the other 4\% of the predicted $\overline{\nu}_e$ flux above 1.8~MeV.  
In the 5-7 MeV range of antineutrino energy, the fractional contributions of these different groups are 97\% and 3\%, respectively.  
These fractional contributions are roughly similar for the fission isotopes $^{239}$Pu and $^{241}$Pu.

\begin{figure}[htb!pb]
\includegraphics[trim=4.2cm 0.6cm 4.5cm 2.6cm, clip=true, width=0.90\linewidth]
{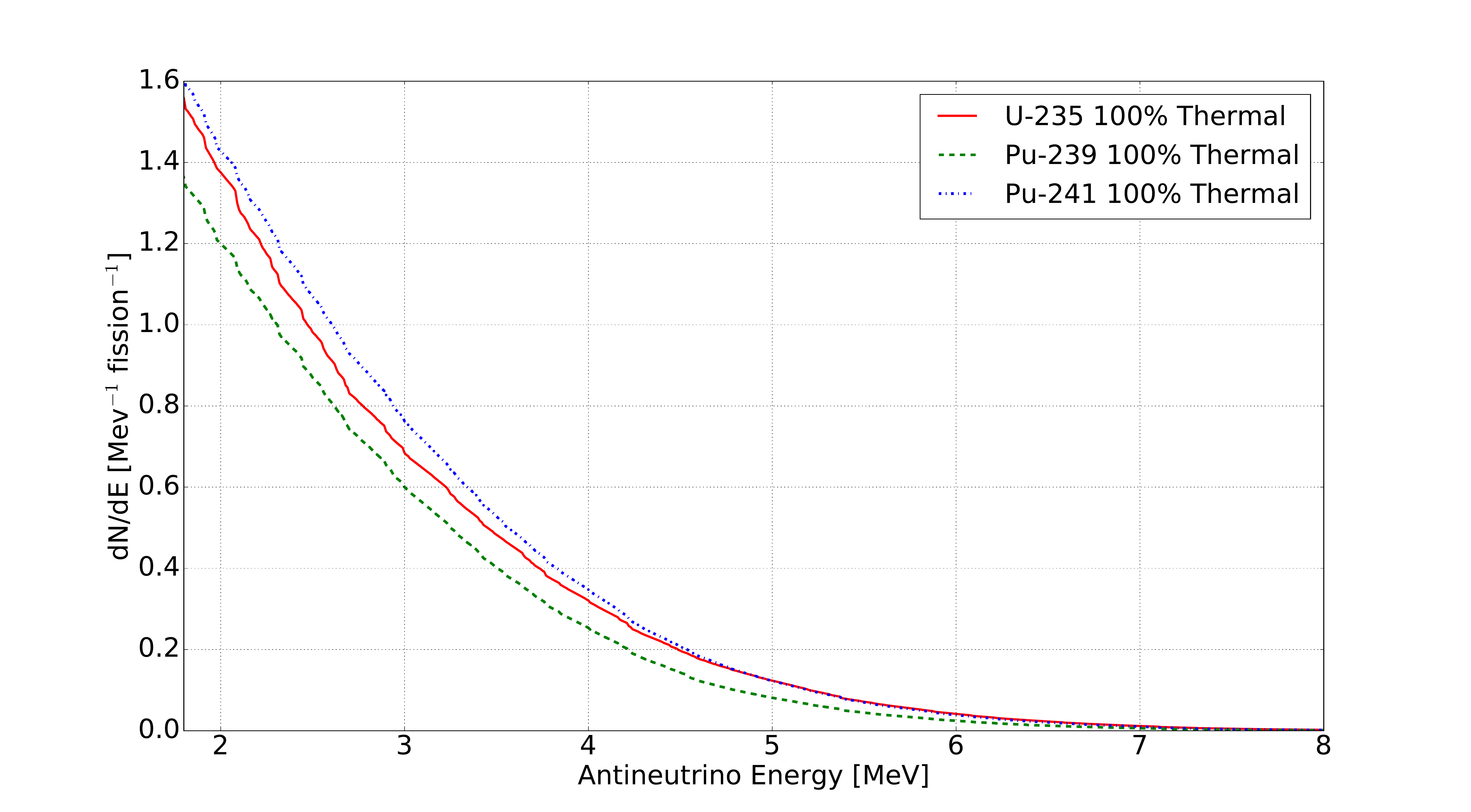}
\caption{Predicted antineutrino spectra per thermal fission of $^{235}$U, $^{239}$Pu, and $^{241}$Pu\@.  Spectra are calculated above the 1.8~MeV threshold for inverse beta decay.  
Pictured predictions use the JEFF-3.1.1 thermal fission yield database, with the default summation treatment, as described in the text.}
\label{fig:Spec}
\end{figure}

The estimated $\overline{\nu}_e$ spectra from thermal fission of $^{235}$U, $^{239}$Pu, and $^{241}$Pu are pictured in Figure~\ref{fig:Spec}.  
It is useful to note some general features of these absolute spectra before considering flux and spectrum differences between neutron energy cases.  
The summation-predicted spectra contain familiar features also found in beta-conversion predictions, such as long high-energy tails and a suppressed flux for $^{239}$Pu relative to $^{235}$U and $^{241}$Pu.  
They also contain unique features noted in previous summation calculations, such as kinks in the spectrum caused by the endpoints of prominent beta branches~\cite{bib:dwyer, sonzogni_fine}.  

\subsection{Impacts on Antineutrino Spectra}

To quantify impacts of neutron energy on the reactor $\overline{\nu}_e$ spectrum, we will compare two neutron energy cases: one in which all fissions are thermal, as in the ILL beamline, and one in which fission energies match those of the LEU pressurized water reactor case in Table~\ref{tab:fissionenergy}, as would have been observed by recent $\theta_{13}$ experiments.  
We will refer to these as the `thermal' and `LEU' cases, respectively.  
It should again be noted that both resonant and fast fission contributions in the LEU case use cumulative fission yields for fast neutrons ($Y^c_f$), due to the lack of resonant information in  fission yield databases.

\begin{figure}[htb!pb]
	\includegraphics[trim=3.5cm 0.7cm 14.0cm 1.8cm, clip=true, width=0.99\linewidth]{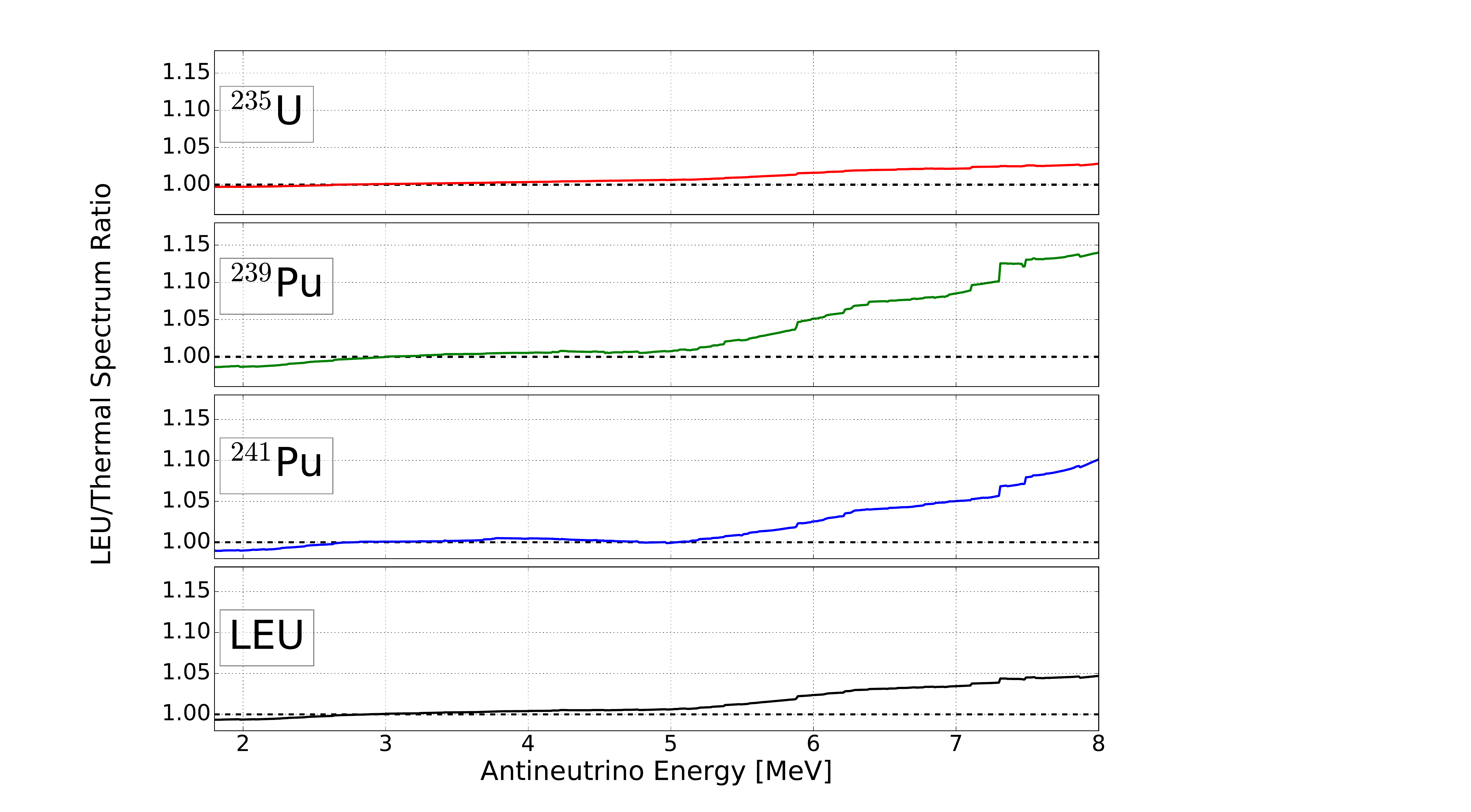}
\caption{Reactor antineutrino energy spectrum ratios between summation $\overline{\nu}_e$ models including fission neutron energies similar to a LEU pressurized water reactor and similar to the ILL thermal neutron beamline.  Ratios for the default calculation described in the text are pictured for $^{235}$U, $^{239}$Pu, and $^{241}$Pu.  LEU/thermal ratios utilizing the Daya Bay-reported fission fractions reported in Ref.~\cite{bib:prl_rate} are shown in the bottom panel; in both LEU and thermal cases, $^{238}$U spectrum contributions are held constant, while contributions for the other isotopes are adjusted as indicated in the top three panels.}
\label{fig:SpecRatio}
\end{figure}

Figure~\ref{fig:SpecRatio} shows the ratio of the LEU to thermal $\overline{\nu}_e$ spectrum per fission  for the three relevant fission isotopes.  
For the largest contributor to the LEU flux, $^{235}$U, this ratio is within $\sim$2.5\% of unity across the entire energy spectrum, with no bump-like features visible in the 5-7 MeV energy 'spectrum anomaly' regime.  
These deviations from unity are smaller in magnitude than existing quoted uncertainties in the beta-conversion $^{235}$U spectrum reported by Huber~\cite{bib:huber}.  

For the plutonium isotopes, similar trends in energy are greater in magnitude, due to the larger contribution of non-thermal fissions in the LEU case.  
While no bump-like feature is present in the 5-7 MeV energy region for the $^{239}$Pu and $^{241}$Pu LEU/thermal ratios, deviations from unity as large as 8\% are visible in this energy range.  
However, as was the case for $^{235}$U, these deviations from unity are still smaller than the quoted beta-conversion prediction uncertainties, which are substantially larger for $^{239}$Pu and $^{241}$Pu.
Thus, proper inclusion of non-thermal fissions in the prediction of a reactor's $\overline{\nu}_e$ spectrum appears unlikely to produce new spectral features extending beyond the bounds of the existing beta-conversion model uncertainty.

This last point is more directly illustrated in the bottom panel of Figure~\ref{fig:SpecRatio}.  In this figure, summation calculations are conducted for a reactor with PWR-like~\cite{bib:cpc_reactor} fission fractions for the dominant isotopes $^{235}$U, $^{238}$U, $^{239}$Pu, and $^{241}$Pu: 0.561, 0.076, 0.307, and 0.056, respectively.  
As before, a ratio is made between the case of purely thermal $^{235}$U, $^{239}$Pu, and $^{241}$Pu fission products (as is effectively realized in existing beta-conversion predictions), and the case where the thermal and non-thermal fission products actually expected in a PWR reactor are considered.  
In both cases, $^{238}$U contributions are identical, as they are assumed to be from fast fissions only.  
This figure provides an estimate of how much the present beta-conversion prediction for LEU cores similar to Daya Bay's might be expected to change if existing beta spectrum measurements instead contained truly representative fission neutron energies.  
As was the case for the individual isotopes, no bump-like feature is present in this ratio.  
Overall deviations from the pure-thermal spectrum prediction are no greater than 3.5\% in the 5-7 MeV energy range.  
Thus, it appears that improper treatment of fission neutron energies is unlikely to be the source of the reactor spectrum anomaly.  

\subsection{Impacts on IBD Yields}

The clustering of all LEU/thermal spectrum ratios around unity provides a good indication that thermal neutron energy treatment has little impact on overall IBD yields for $^{235}$U, $^{239}$Pu, and $^{241}$Pu.  
This observation is more explicitly demonstrated in Table~\ref{tab:IBDratios}, which shows IBD yield per fission ratios between the LEU and thermal cases.  
We note that IBD yields are obtained by multiplying the $\overline{\nu}_e$ spectra described above by the inverse beta decay interaction cross-section.  
Differences in IBD yields between the two cases are found to be less than 1\% for all three  isotopes.  
This is significantly smaller than the existing normalization uncertainties associated with the beta conversion process.   

\begin{table}[thpb!]
\caption{IBD yield ratios between summation $\overline{\nu}_e$ models including fission neutron energies similar to a LEU pressurized water reactor and similar to the ILL thermal neutron beamline.  Ratios for the default calculation described in the text are pictured for $^{235}$U, $^{239}$Pu, $^{241}$Pu, and for an LEU core with Daya Bay's reported fission fractions, as in Figure~\ref{fig:SpecRatio}.  LEU/thermal ratios are also provided for alternate summation calculations that use different inputs as described in Section~\ref{sec:Syst}.}
\begin{tabular}{c|c|c|c|c}
\hline \hline
\multirow{2}{*}{Scenario} &  \multicolumn{4}{|c}{LEU/Thermal IBD Yield Ratio} \\ \cline{2-5}
& $^{235}$U & $^{239}$Pu & $^{241}$Pu & LEU\\ \hline \hline
JEFF, Default Isotopes & 1.001 & 0.997 & 0.998 & 0.999 \\ \hline \hline
ENDF, Default Isotopes & 1.003 & 1.000 & 0.998 & 1.001 \\ \hline
JEFF, All Isotopes & 1.001 & 0.999 & 1.000 & 1.000 \\ 
\hline \hline
\end{tabular}
\label{tab:IBDratios}
\end{table}


Similar to the last section, we consider the LEU/thermal IBD yield ratio for a reactor exhibiting PWR-like fission fractions, which is also given in Table~\ref{tab:IBDratios}.
Unsurprisingly, IBD yields for the pure-thermal and LEU-like fission neutron energy cases are found to differ by substantially less than 1\%.  
This observation indicates that fission neutron energy differences are likely not responsible for the reactor antineutrino flux anomaly.  
Moreover, since no specific isotope deviates substantially from unity in this ratio, fission neutron energy treatment is also unlikely to explain the differences in flux evolution between beta-conversion models and that observed recently at Daya Bay.  

\subsection{Impacts on Comparisons Among Reactor Experiments}

In the preceding sections, we drew comparisons between the thermal case, which includes a 0\% non-thermal fission contribution, and the LEU case, which includes an 18\% non-thermal contribution for $^{235}$U.   
For HEU experiments, such as PROSPECT at the HFIR reactor~\cite{prospect}, this non-thermal contribution is 14\%, as indicated in Table~\ref{tab:fissionenergy}, quite similar to the LEU case.  
In fact, for the LEU and HFIR $^{235}$U cases, both the resonant (15\% and 12\%, respectively) and fast (2\% for both) fission contributions are quite consistent.  
The similarity in non-thermal contributions for $^{235}$U indicates that differences in fission neutron energies between HEU and LEU experiments may result in at most $\mathcal{O}$(0.1\%) biases in comparisons of HEU and LEU IBD yields and spectra.  
This is negligible compared to the variety of other systematic uncertainties that would be encountered in such a comparison, such as energy scale uncertainties in IBD spectrum comparisons, or detection efficiency and reactor thermal power uncertainties in IBD yield comparisons. 
Comparisons of rates and spectra among HEU experiments are also likely to be unaffected by fission neutron energy differences.  
In Table~\ref{tab:fissionenergy}, we see $^{235}$U non-thermal contributions of 6\% and 14\% for the two HEU reactors.  
Scaling from the $^{235}$U case pictured in Fig.~\ref{fig:SpecRatio} (0\% versus 18\% non-thermal),  spectrum deviations between these two HEU cases should be at most around the 1\% level, once again negligible compared to other expected detector and model systematics.  

It may be possible to explicitly measure IBD yield and spectrum variations resulting from neutron energy differences by comparing current experiments at HEU cores to future potential experiments at other types of reactors.  
A future short-baseline experiment at a highly $^{235}$U-enriched advanced fast reactor (AFR) core, where nearly all fission neutrons are non-thermal, would contain an $\sim$82\% difference in non-thermal contribution with respect to PROSPECT at HFIR.  
Scaling from the $^{235}$U case pictured in Fig.~\ref{fig:SpecRatio}, this large difference in non-thermal fissions could possibly induce measurable $\mathcal{O}$(10\%)-level variations in the IBD spectrum at high energies.  

\section{Result Cross-Checks}
\label{sec:Syst}

Thus far, systematic uncertainties in the summation calculation have not been considered.  We use the following section to examine the impact of limitations in database-reported fission yields, in nuclear structure data for some fission products, and in the use of fast fission yields for the resonant fission case.  

All fission yield measurements summarized in the JEFF database are accompanied by sizable systematic uncertainties.  
Given the difficulty in determining a proper treatment of the correlations among fission yield database entries, we did not attempt the full propagation of these uncertainties in our summation calculations.  
As an alternative, we compare JEFF's reported fission yields to those reported in other standard nuclear databases.  
In Table~\ref{tab:ratios}, we list the ten largest contributors to the $\overline{\nu}_e$ flux in the 5-7~MeV range for $^{235}$U and $^{239}$Pu, along with their JEFF-reported thermal fission yields and JEFF- and ENDF-reported thermal-fast yield differences, $Y_t-Y_f$.  
It should be noted that the ENDF fission yield database contains deficiencies for a number of crucial isotopes~\cite{sonzongi2}; we find that many isotopes identified in Ref.~\cite{sonzongi2} containing deficient treatment of isomeric ratios in the ENDF~$^{235}$U thermal database show similar deficiencies for ENDF~$^{235}$U fast fission yields and $^{239}$Pu fission yields.  
For these isotopes, JEFF fission yields are used in place of ENDF-reported yields.  
Prominent affected/altered isotopes are highlighted in Table~\ref{tab:ratios}.

\begin{table}[hptb!]
\caption{Thermal fission yields $Y^c_t$ and thermal-fast yield differences, $Y^c_t-Y^c_f$, for isotopes with the largest contribution to the $^{235}$U and $^{239}$Pu 5-7 MeV antineutrino flux.  Values of $Y^c_t-Y^c_f$ are provided for the JEFF and ENDF fission yield databases, as well as Q-value and $N$, the relative flux contribution to the 5-7 MeV range of antineutrino energy, in percent.  A `*' denotes a metastable state for that isotope, while a `\^~' indicates that JEFF fission yield values are used in place of ENDF fission yield values, for reasons described in the text.}  
\begin{tabular}{c|c|c|c|c|c}
\hline \hline
\multirow{2}{*}{Isotope} & $Y^c_t$ & $Y^c_t-Y^c_f$ & $Y^c_t-Y^c_f$ & N(5-7) & Q-Value  \\ 
 & (JEFF) & (JEFF) & (ENDF) & (\%) & (MeV) \\ \hline
\multicolumn{6}{c}{$^{235}$U} \\ \hline 
Y-96 &  0.047 & -0.0004 & -0.0004\^ & 10.66 & 7.10 \\
Rb-92 & 0.048 & -0.0032 & +0.0064 & 9.63 & 8.10 \\
Cs-142 & 0.029 & -0.0025 & -0.0012 & 5.77 & 7.32 \\
Nb-100 & 0.056 & - 0.0036 & -0.0003 & 4.61 & 6.38 \\
Rb-93 & 0.035 & -0.0064 & -0.0021 & 3.92 & 7.47 \\ 
Cs-140 & 0.060 & +0.0034 & -0.0002 & 3.26 & 6.22 \\
I-138 & 0.015 & +0.0009 & +0.0013 & 3.09 & 7.99 \\
Y-99 & 0.019 & -0.0103 & -0.0038 & 3.05 & 6.97 \\ 
Rb-90 & 0.044 & +0.0051 & +0.0023 & 3.03 & 6.58 \\
Sr-95 & 0.053 & -0.0004 & +0.0003 & 3.01 & 6.09 \\ 
\hline 
\multicolumn{6}{c}{$^{239}$Pu} \\ \hline 
Y-96 &  0.029 & -0.0015 & -0.0015\^ & 10.86 & 7.10 \\
Nb-100 & 0.052 & +1.6e-5 & +1.6e-5\^ & 7.16 & 6.38 \\
Nb-102* & 0.016 & -0.0039 & -0.0039\^ & 6.85 & 7.26 \\
Rb-92 & 0.020 & -0.0035 & -0.0009 & 6.73 & 8.10 \\
Cs-142 & 0.016 & +0.0043 & +0.0019 & 5.35 & 7.32 \\
Cs-140 & 0.044 & +0.0026 & -0.0047 & 4.02 & 6.22 \\
Y-99 & 0.013 & -0.0045 & +0.0017 & 3.60 & 6.97 \\ 
Rb-93 & 0.017 & -0.0050 & -0.0015 & 3.11 & 7.47 \\ 
Y-98* & 0.019 & -0.0051 & +0.0014 & 3.08 & 9.40 \\ 
Sr-95 & 0.032 & -0.0003 & -0.0021 & 3.07 & 6.09 \\ 
\hline \hline
\end{tabular}
\label{tab:ratios}
\end{table}

\begin{figure}[htb!pb]
	\includegraphics[trim=3.5cm 0.5cm 5.5cm 2.8cm, clip=true, width=0.95\linewidth]{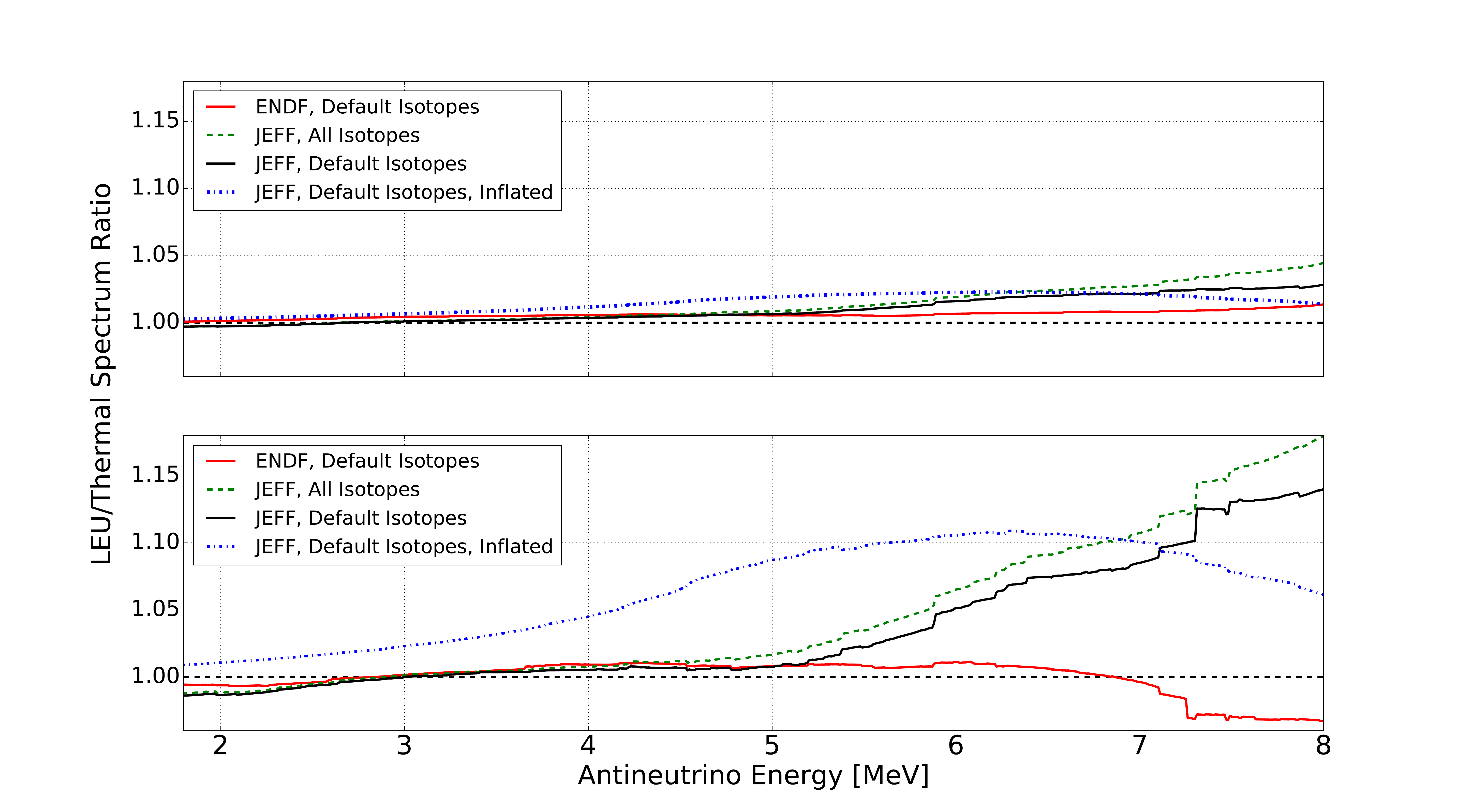}
\caption{Reactor antineutrino energy spectrum ratios between summation $\overline{\nu}_e$ models including fission neutron energies similar to a LEU pressurized water reactor and similar to the ILL thermal neutron beamline.  Ratios for the default cases described in the text are pictured for $^{235}$U (top) and $^{239}$Pu (bottom), along with ratios for a variety of other cases described in the text.}
\label{fig:SpecRatioSyst}
\end{figure}

Differences in ENDF- and JEFF-reported fission yields should naturally lead to differences in reported $\overline{\nu}_e$ fluxes and spectra for each fissioning isotope.  
Figure~\ref{fig:SpecRatioSyst} shows $^{235}$U and $^{239}$Pu LEU/thermal spectrum ratios for the default (JEFF) case described above, as well as the case in which ENDF yields are used instead.  
While the JEFF yields produce a positive ratio at high energies, the ENDF case is relatively flat across the spectrum, indicating little to no change in the $\overline{\nu}_e$ spectrum between the LEU and thermal cases.
The comparative flatness of the ENDF-derived spectrum ratio reflects differences seen in $Y^c_t-Y^c_f$ values, which are somewhat visible in Table~\ref{tab:ratios}: the JEFF database favors higher fast fission yields for isotopes contributing the most high-energy neutrino flux (a negative $Y^c_t-Y^c_f$), while ENDF is slightly more evenly split between fast and thermal cases.  
In Table~\ref{tab:IBDratios}, we also provide LEU/thermal IBD yield ratios derived from using the ENDF fission yields; as in the JEFF case, deviations from unity are $<$1\%.  
Thus, when using corrected ENDF fission yield data, we obtain conclusions largely identical to those in the previous section.  

The conclusions of the default case also rely on the assumption that fission yields are roughly equivalent for fast and resonant fissions of the same isotope.  
The completeness and precision of existing resonant fission yield measurements are limited, and the data do not clearly point to a more reasonable approach.  
While this assumption is almost certainly incorrect, it should provide a conservative overestimate of the impact from non-thermal fission.
To cross-check the results of this approach, we also calculate LEU/thermal spectrum and flux ratios for an alternate treatment of resonant fission yields.  
In Ref.~\cite{hayes2}, it is noted that existing data on resonant fission yields for high-yield isotopes do not rule out changes on the order of 20\% with respect to thermal yields.  
Thus, we examine the extreme case in which the 20 largest contributors to the 5-7 MeV $\overline{\nu}_e$ flux for thermal $^{235}$U and $^{239}$Pu have resonant fission yields systematically higher than the thermal case by 20\%; for all other isotopes, thermal and resonant fissions are assumed to be identical.  
While this assumption is likely unrealistic (as indicated by the aggregate epithermal fission yield measurements cited above~\cite{Leconte2012}) it provides an additional worst-case test (allowed by existing yield data) with which to bracket the possible impact of non-thermal fissions on the $\overline{\nu}_e$ spectrum observed at reactors.  
We note that since this inflation does not conserve the overall number of fission daughters per fission, we do not consider IBD yield variations produced by this scenario.  
Spectral LEU/thermal ratios for this scenario are also shown in Figure~\ref{fig:SpecRatioSyst}, labeled as the 'Inflated' case.  
Even when all top-producing isotopes are given inflated resonant fission yields, deviations from unity in this ratio are at most 2.5\% for $^{235}$U, and 11\% for $^{239}$Pu.  
Given the dominance of $^{235}$U in the overall spectrum, these deviations are not sufficient to produce the observed $~$10\% bump in the spectrum at 5-7~MeV.  
Thus, this extreme approach also provides an identical conclusion to that described in the previous section.  

Beyond the limitations in fission yields, knowledge of the nuclear structure of many fission  daughters is severely limited.  
Many isotopes with fission yields specified in the JEFF and ENDF databases contain incomplete descriptions of daughter excited states and beta feedings $f_{ij}$, making a calculation of a $\overline{\nu}_e$ spectrum for that isotope impossible.  
Thus, in the default case, these isotopes are excluded from the summation calculation.  
To assess the possible impact of these isotopes on the conclusions of the previous section, we take the default summation calculation and add the simplistic assumption that all isotopes with incomplete decay information feed entirely to the ground state.  
 While this assumption cannot possibly be true, it serves to maximize contributions to the the high-energy region of the $\overline{\nu}_e$ flux and to any relative differences between spectral features in the thermal and fast cases.  
 We find that including these isotopes in this manner increases the overall $\overline{\nu}_e$ flux from thermal $^{235}$U fissions by 6.1\%, and increases the contribution in the 5-7 MeV energy region by 7.8\%; these flux increases are roughly similar for $^{239}$Pu and for fast fission cases.  
 
The change to LEU/thermal ratios arising from this flux increase is shown in Figure~\ref{fig:SpecRatioSyst} and in Table~\ref{tab:IBDratios}, labeled as the `All Isotopes' case.  The deviation from unity of the LEU/thermal IBD spectrum ratio is found to be enhanced with respect to the default case by $<$1\% in the 5-7 MeV range for $^{235}$U, and by an additional 1-2\% for $^{239}$Pu.  
IBD yield ratios are within 0.2\% of the default case for all three isotopes.  
Thus, it appears that while a simplistic addition of the remaining $\overline{\nu}_e$ flux from poorly-understood beta branches does increase the difference between thermal and non-thermal $\overline{\nu}_e$ spectra, it does not substantially alter the conclusions drawn from the default case.

\section{Summary}
\label{sec:Implications}

In this study, we used summation calculations to estimate the role of fission neutron energy in $\overline{\nu}_e$ production via fission of $^{235}$U, $^{239}$Pu, and $^{241}$Pu.  
Using reactor core simulations, we first examined what variations in fission neutron energies exist among reactors suitable for $\overline{\nu}_e$ measurements.  
Non-thermal neutrons were found to be responsible for 6-18\% of all $^{235}$U fissions for the different LEU and HEU core types considered.  
In contrast, for $^{239}$Pu, and $^{241}$Pu, which account for a substantial fraction of fissions in PWR reactors, most fissions were non-thermal: 70\% and 62\% of all fissions are non thermal for $^{239}$Pu, and $^{241}$Pu, respectively.  
This high fraction of non-thermal fissions is primarily the result of the presence of a large low-energy fission cross-section resonance present for these two isotopes.  
Current beta-conversion $\overline{\nu}_e$ predictions, which use beta spectrum measurements from thermal neutron beamlines, do not account for the possible $\overline{\nu}_e$ spectrum and flux deviations arising from these substantial non-thermal fission contributions.  

We then used the Oklo nuclear modeling toolkit to generate summation $\overline{\nu}_e$ predictions for fissions from purely thermal neutrons and from neutrons matching the significantly non-thermal energy profile described in the previous paragraph.  
Antineutrino fluxes and spectra were then compared between the two cases. 
For $^{235}$U, $^{239}$Pu, and $^{241}$Pu, we find that differences in $\overline{\nu}_e$ spectrum and IBD yields between these cases are smaller than the corresponding beta-conversion uncertainty envelope for each isotope.  
Of these three isotopes, $^{239}$Pu~and $^{241}$Pu exhibit the largest spectrum deviations due to the large non-thermal fission contributions of these isotopes.  
When considering a reactor with PWR-like fission fractions, the difference in the $\overline{\nu}_e$ spectrum between these two cases is less than 3.5\% in the 5-7 MeV region and below, and the difference in IBD yield is much less than 1\%.  
This result indicates that fission neutron energy treatment is unlikely to be responsible for the reactor antineutrino flux and spectrum anomalies, or for the difference in flux evolution between Daya Bay observations and beta-conversion predictions.  
These conclusions appear to be consistent regardless of which fission yield database is used or how incomplete nuclear structure information is treated.  

We also find that the considered HEU and LEU reactor cores exhibit comparatively similar $^{235}$U spectra despite differences in non-thermal fission contributions.  
Thus, while our study was not inclusive of all so-called 'thermal' core types, it seems quite likely that fission neutron energy differences can be safely neglected in IBD yield and energy spectrum comparisons between most HEU and LEU reactor $\overline{\nu}_e$ experiments, or among experiments at differing HEU or differing LEU reactors.  
This observation is relevant to joint sterile oscillation analyses between reactor experiments~\cite{bib:JointAnal}, to joint HEU-LEU efforts aimed at identifying the isotopic origin of the reactor flux and spectrum anomalies~\cite{surukuchi}, and to global fits to reactor flux and spectrum measurements~\cite{GiuntiMe,JointSchwetz,GiuntiRatio}.  

A variety of future experiments could elucidate the relative differences in $\overline{\nu}_e$ production by fast, resonant, and thermal fissions.  
Additional measurements of fission yields at specific non-thermal energies for individual Pu daughters with dominant contributions to the high-energy $\overline{\nu}_e$ flux could help bracket non-thermal fission effects in summation $\overline{\nu}_e$ calculations.  
As previously mentioned, these differences could be probed directly for $^{235}$U via future $\overline{\nu}_e$ measurements at fast HEU reactors.  
Existing reactor designs are not likely to enable similar measurements for $^{239}$Pu and $^{241}$Pu.  
For these isotopes, dedicated measurements of fission beta spectra similar to those made at ILL for a variety of non-thermal energies would be useful.  
Such measurements, particularly in the vicinity of both isotopes' $\sim$0.3~eV fission cross-section resonance, will produce beta-conversion $\overline{\nu}_e$ predictions more representative of the true $\overline{\nu}_e$ spectrum produced in low-enriched uranium reactor cores than the existing ILL measurements.  





\section{Acknowledgement}
We thank A. Hayes and A. Sonzogni for providing nuclear structure and fission yield data files used in this work.  
The work of the IIT group was funded by DOE Office of Science, under DOE OHEP DE-SC0008347, as well as by the IIT College of Science.  
The research of the Georgia Tech group was performed under appointment to the Nuclear Nonproliferation International Safeguards Fellowship Program sponsored by the Nation Nuclear Security Administration’s Office of International Nuclear Safeguards (NA-241).
Work at Lawrence Berkeley National Laboratory was supported under DOE OHEP DE-AC02-05CH11231.
\bibliographystyle{apsrev4-1}
\bibliography{main}{}

\end{document}